\documentclass{aa}
\usepackage[varg]{txfonts}
\bibpunct{(}{)}{;}{a}{}{,}
\usepackage{graphicx}
\usepackage{algorithm}
\usepackage[noend]{algpseudocode}
\usepackage{cellspace}
\usepackage{url}
\usepackage{color}
\usepackage[toc, page]{appendix}
\usepackage[utf8]{inputenc} %unicode support

\begin{document}

\title{Identifying Earth-impacting asteroids using an artificial neural network\\ }
\author{John D. Hefele\inst{1}
\and Francesco Bortolussi\inst{2}
\and Simon Portegies Zwart\inst{1}}
\institute{Sterrewacht, Leiden University, Leiden, NL
\and LIACS, Leiden University, Leiden, NL}
\date{Received 29 May 2019 / Accepted 9 December 2019}

\abstract{By means of a fully connected artificial neural network, we identified asteroids with the potential to impact Earth. The resulting
instrument, named the Hazardous Object Identifier (HOI), was trained on the basis of an artificial set of known impactors which were
generated by launching objects from Earth's surface and integrating
them backward in time. HOI was able to identify
95.25\% of the known impactors simulated that were present in the test set as potential impactors. In addition, HOI was able to identify 90.99\% of the potentially hazardous objects identified by NASA, without being trained on them directly.}


\titlerunning{On the identification of
  Earth-impacting asteroids using an artificial neural network}

\authorrunning{J. D. Hefele et al.}
\maketitle

\section{Introduction}
\label{SEC:Introduction}

In 1990 the US Congress requested for NASA to establish two workshops
to focus on the identification of potentially hazardous small bodies
and on methods of altering their orbits to prevent impact
\citep{milani2002}. The workshops led to the establishment of the
\textit{Sentry earth impact monitoring} system \citep{Sentry}.  If a
hazardous asteroid is identified early enough prior to
impact, it would be possible to mitigate the impact by means of an
appropriate space mission to alter the asteroid's orbit through a
gravitational tugboat \citep{10.2307/26060526} or by obliterating it
with a nuclear warhead \citep{BARBEE201837}.  Both mitigation
strategies require many years of preparation, which makes the early
detection of hazardous objects vital for allowing ample time to prepare such
missions.

The Sentry system adopts a Monte Carlo approach in which millions of
virtual objects are launched with orbital parameters that are statistically
sampled from within the error ellipse of the observed asteroids. The
impact probability is subsequently determined based on the fraction of
virtual asteroids that reach Earth within some predetermined striking
distance \citep{milani2002}. In this approach, the orbits of many
asteroids are integrated numerically and the final parameter space is
considered to represent the probability-density distribution of the
respective objects. The calculation of this probability density
distribution relies on the algorithm and implementation used to
integrate the orbits of the asteroids.  The time scale over which such
integrations remain reliable depends on the degree to which the
asteroid's orbit is chaotic, that is, it depends on the value of the largest positive
Lyapunov exponent.  Additionally, the reliability of such integrations
depends on the ability of the integrator to obtain a solution, such that the integration complies to the concept of nagh Hoch\footnote{Nagh Hoch is a concept stating that an ensemble of random initial realizations in a wide range of parameters gives statistically the same result as the converged solutions of the same ensemble of realizations.}
\citep{PORTEGIESZWART2018160}.
\par
Both of these concepts are not
guaranteed with regard to the adopted numerical schemes and the results reach
questionable proportions as soon as the asteroid experiences a close
encounter with any object other than the Earth. In the latter case,
the phase space of possible solutions grows exponentially due to the
chaotic nature of the equations of motion.  Establishing the chaotic
nature of an asteroid is limited by the accuracy of its orbital
determination. This is generally realized by observing any particular
asteroid a number of times. These observations result in a data arc,
the fraction of the orbit over which the object has been observed.
The adopted Monte-Carlo method used in the Sentry system is expected
to be reliable for at most a few dozen years
\citep{HorizonsManual} for asteroids whose observed data arc
is shorter than a month, which comprises 12.9\% of all smallbodies \citep{dastcom5}.

Considering the high degree of chaotic motion (small Lyapunov time
scale) in asteroids and the consequential exponential divergence of
its orbit, one might wonder if it is worth the effort to perform
extensive computer simulations to track the orbital
trajectories of a large number of particles so long as the veracity of
the orbital integration cannot be guaranteed. For the most chaotic
asteroids, the impact probability depends acutely on the
statistics of the adopted method and a more coarse grained approach to
identify potentially hazardous objects may suffice. This approach
would free up computer time to provide a more reliable impact probability for the most promising candidate impostors. 

We explore the population of asteroids and, in
particular, the potentially dangerous ones by means of automatic
machine recognition through a combination of numerical integrations and
a trained neural network similar to the architectures described in \citet{ref1} and \citet{ref2}, which were used for classifying hazardous taxonomy and solar sail transfer time estimation respectively. It is a statistical approach in which we
determine the prospect for impact of the known population of asteroids
gathered from the \textit{dastcom5} off-line database \citep{dastcom5}.  Our
analysis is mediated by an artificial neural-network dubbed HOI\footnote{This also means ``Hello'' in the Dutch language.} for Hazardous Object Identifier, which was trained
on a population of known impactors (KI) and a random sample from the
observed database using the \textit{TensorFlow} framework \citep{TensorFlow}. The KIs are machine-generated from an integrated
population of asteroids that start their orbit on a random position of
Earth's surface and are launched radially away with the varying speeds. These objects are subsequently integrated backward in time
together with the planets in the Solar System for up to 20,000
years. To train HOI, these computer generated KIs are then mixed with a subset of observed asteroids, which we assume to be known non-impacting objects. The trained network is then used on another random selection of observed asteroids
in order to identify potential impactors (PIs). All the objects that were not identified by the model as PIs, which were not initially labeled as KIs, are referred to as unidentified objects (UOs).

We begin by describing HOI's architecture in Section\,\ref{SEC:arch}, followed by a discussion of the generation of the small-body datasets in Section\,\ref{SEC:Data}. The results are examined in Section\,\ref{SEC:Results} and conclusions are drawn in Section\,\ref{SEC:Conclusions}. All the code
used to train the neural network, generate data, and evaluate the
results are publicly available on GitHub\footnote{ \url{https://github.com/mrteetoe/HOI}}.
%\citep{koon,oreg,khar,zvai,xjon,schn,pond,smith,marg,hunn,advi,koha,mouse}

\section{Hazardous Object Identifier (HOI)}
\label{SEC:arch}

In general, neural networks are particularly well-suited for
recognizing complex patterns hidden in multidimensional datasets. In our particular case, we strive to identify observed objects that have topologically similar trajectories to the trajectories of the population of KIs. Because we are no longer reliant on calculations that attempt to estimate the asteroids position at a particular point in time, the network is more resilient to perturbations of the initial conditions, that is, chaotic motion.

The problem at hand is a discrete binary classification task, where the two mutually exclusive classes for the observed objects are either potential impactors (PIs) or unidentified objects (UOs). For the purpose of our experiments, the UOs are what we would consider ``benign objects'', meaning objects that are identified as having a negligible chance of colliding with the Earth. To quantify the network's accuracy, the standard cross-entropy cost function is used. This is defined as:
\begin{equation}
        \label{cost_function}
        H(y,\hat{y})=-\sum_i^{N} y_i \text{ln}(\hat{y}_i)+(1-y_i)\text{ln}(1-\hat{y}_i).
\end{equation}
Here $y$ is the actual value, or label, $\hat{y}$ is the predicted value, and $N$ is the total amount of predictions. This cost function has the convenient property that its derivative with respect to some input weight, $w$, scales linearly with the difference between the label and predicted value \citep{Nielson}:
\begin{equation}
        \frac{\partial C}{\partial w}=\frac{1}{N}\sum^{N}_i x(\hat{y}_i-y_i)
\end{equation}
Here $x$ is the input value by which $w$ is multiplied. To minimize (\ref{cost_function}), the \textit{Adam Optimizer} is used, which expands upon na\"{i}ve stochastic gradient descent by adapting its learning rate based on both the average of the first and second moments of the gradients \citep{AdamOptimizer}. Empirically, it is observed that this optimizer reduces the cost function to the lowest value with the fewest number of iterations relative to the other algorithms available in TensorFlow. 
\par 
    Each object fed into the HOI is represented by a five-element vector where each vector is the Keplerian elements of the asteroid around the sun including the semi-major axis (\textit{a}), eccentricity (\textit{e}), inclination (\textit{i}), the mean speed (\textit{N}), and the specific angular momentum (\textit{H}).
    These five orbital elements fully characterize the shape of an asteroid trajectory around the sun, but not its orientation as the longitude of the ascending node $\Omega$ and argument of periapsis $\omega$ are omitted.
\par 

\begin{figure}[t]
    \hspace*{-0.4cm}
        \includegraphics[width=83mm]{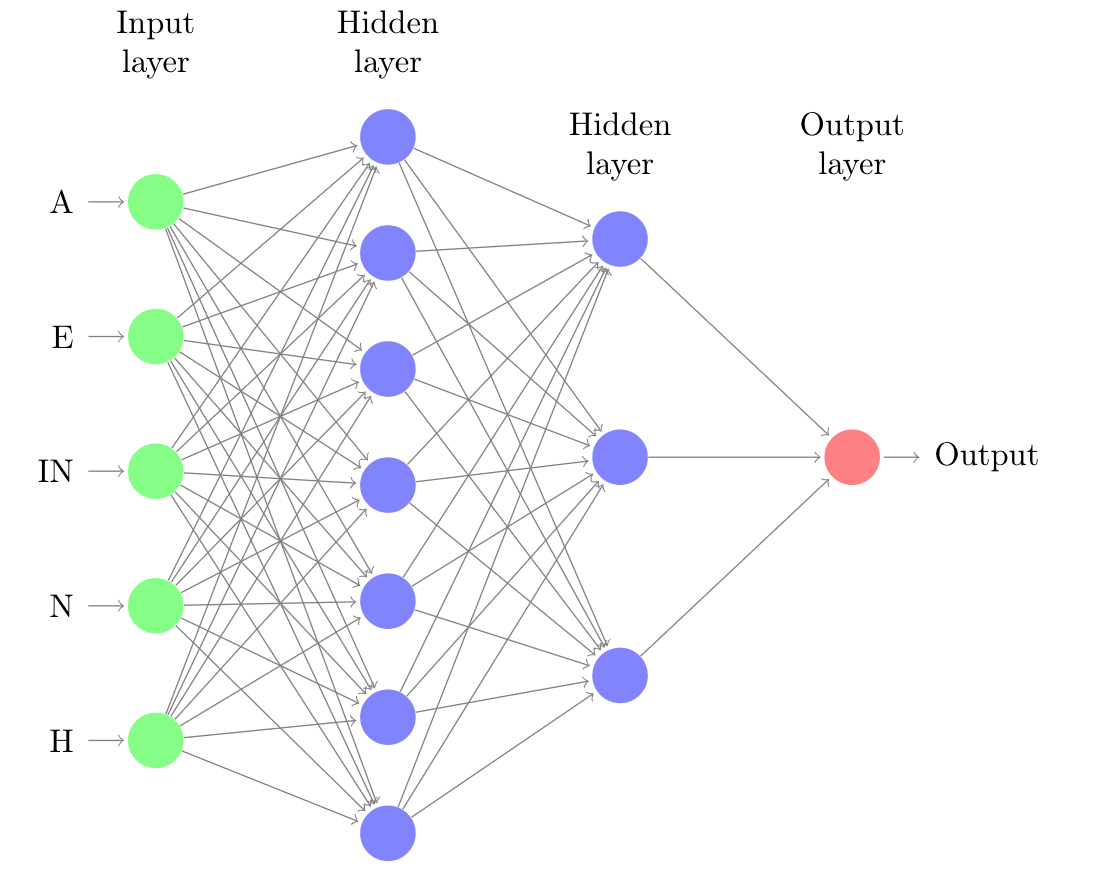}
        \centering
        \caption{\label{FIG:HOI_Design} HOI network architecture. The input layer is comprised of five nodes, which is followed by two hidden layers of seven and three nodes, and an output layer of a single node.}
\end{figure}
A diagram showing the HOI architecture is presented in Fig. \ref{FIG:HOI_Design}. The input layer is a vector of $five$ neurons that matches the dimensionality of the input, which is followed by two hidden layers that are composed of seven and three neurons, respectively, from the input layer. The output layer is composed of a single neuron whose values are restrained between 0 and 1 by virtue of the sigmoid function. Here, objects with a rating of 0.5 or above are classified as PI while those below the threshold are classified as UO.  This neural network architecture was arrived at by a combination of empirical experimentation and the incorporation of domain knowledge. We wanted to provide the network with enough degrees of freedom to properly generalize the orbital elemental profiles of KI but to avoid giving it so many degrees of freedom that the network would overfit to the training datasets.    
\par
\begin{figure}[t]
        \hspace*{-0.3cm}
        \includegraphics[width=83mm]{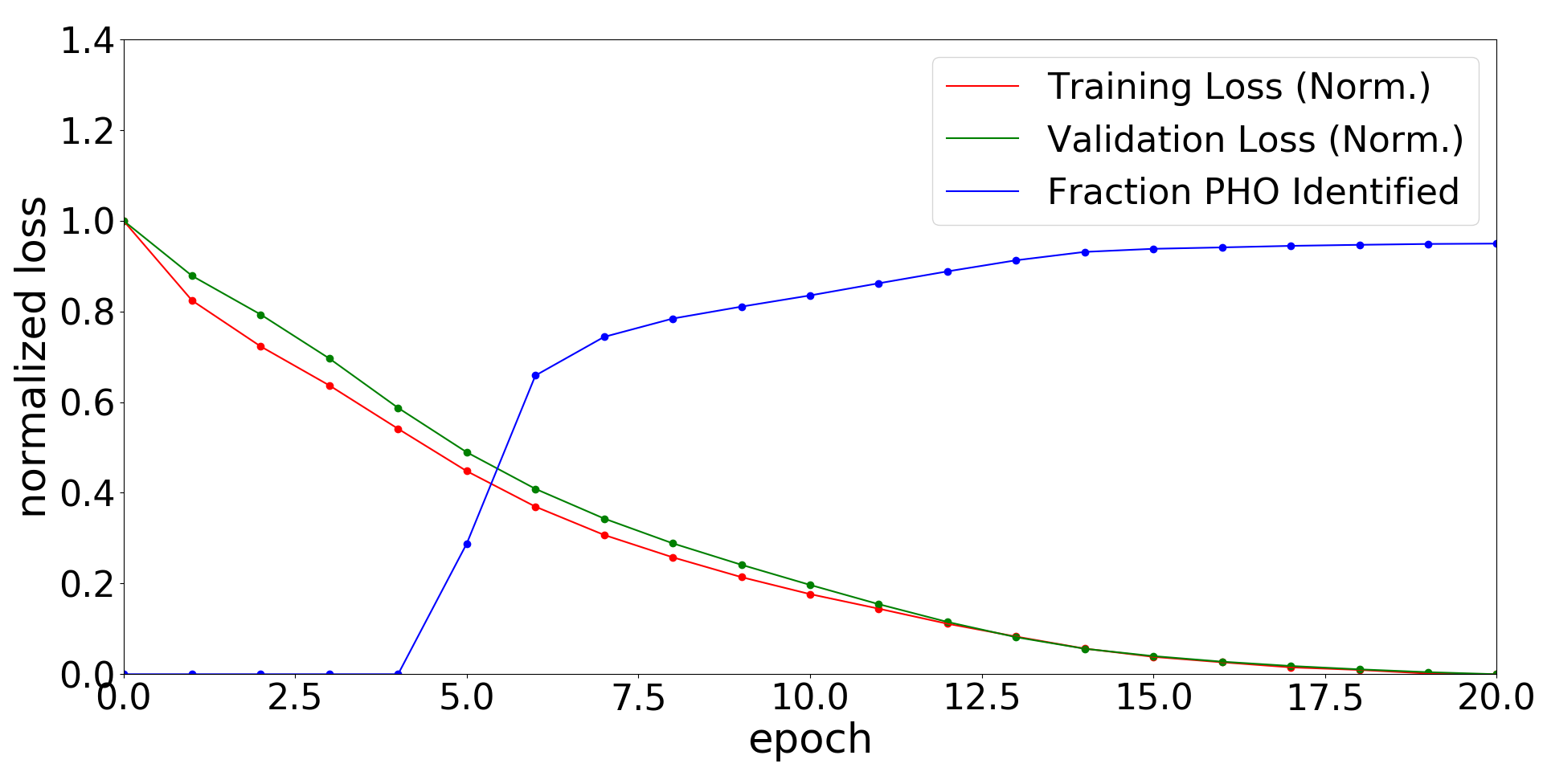}
        \centering
        \caption{\label{FIG:loss} Normalized training and validation losses plotted against the training epoch number, along with the fraction of PHOs identified by the network.}
\end{figure}

  
The described architecture results in 69 free parameters: 59 weights and ten biases \footnote{Following the architecture described, the number of free parameters can be calculated as follows: the input is fed through layers which are comprised of 7, 3, and 1 neuron(s). This results in 5$\times$7+7$\times$3+3$\times$1 weights and 7+3 biases, as only the hidden layers have bias parameters.}. To optimize these parameters, the network is trained on five randomly selected sub-sets of 100,000 observed and KI objects over 20 epochs, which took less than five minutes on a CPU-type laptop without a GPU. The training was halted when the relative loss decrease per epoch was less than $1\%$ to prevent overfitting. At the end of the training process, the network's performance was validated with a subset of 20,000 KI and 20,000 observed objects that had been held out of the training process. Furthermore, all potentially hazardous objects (PHOs)\footnote{All objects with a minimum orbit intersection distance of 0.05 AU or less and an absolute magnitude (H) of 22.0 or less are considered PHOs \citep{NasaPHA}.} were held out of the training process and used exclusively for testing purposes. Fig. \ref{FIG:loss} shows how the training and validation loss decreased per training epoch, while the fraction of PHO hazardous objects identified simultaneously increased.

We gave the observed objects and KIs labels of 0.1
and 0.9, respectively. Here, higher numbers correspond with a larger
probability of colliding with Earth. The label of 0.9 was chosen for the KIs to represent calculations of the KI trajectories which are not converged
solutions \citep{2014ApJ...785L...3P} and  to show that several perturbing effects in
the Solar System were neglected during the simulations, implying that all of the KIs will, in fact, not collide with Earth when their respective velocities are negated. 

To arrive at the label of 0.1 for the observed objects, we assumed that any individual observed object is very likely to be benign by the following logic: first,  all of the PHOs which have considerably larger probability to collide with the Earth compared with the rest of the observed population are not used in HOI training. As a result, their labeling does not degrade the network's ultimate performance. Second, impacts from large objects are rare \citep{Chapman} as the impact frequency of an asteroid collision decreases with the cube of an asteroid's diameter. Earth collisions with 5 kilometer asteroids occur approximately every 20 million years, while those with a 100 meter asteroids occur every 500 years \citep{Bostrom}. Because 98.4\% of the observed objects used for our experiments are greater than 100 meters in diameter\footnote{This assumes an albedo of 0.15 for all small bodies.}, we can use the following formula to estimate an upper-bound of the number of expected Earth impacts from asteroids in our sample within the next 20,000 years:
\begin{equation}
    \label{num_collisions}
    N_{collisions}=\int^{\infty}_{100}\frac{4\times10^7}{D^3}=2000,
\end{equation}
Where $D$ is the diameter of an asteroid. Given that over 700,000 objects were used in HOI training, the number of 2000 mislabeled objects implies that 0.3\% of the observed labels are inaccurate. As discussed further in the following sections, although our sample contains only a small fraction of misclassified non-impactors, they still may effect the ability of HOI to accurately identify an impactor.

\section{Data generation and acquisition}
\label{SEC:Data}

\subsection{Observed objects}

We extracted $736,496$ minor bodies from NASA's \textit{dastcom5}
database \citep{dastcom5}. A percentage of 95.5\% of the extracted objects are
main-belt asteroids, 3.2\% are asteroids that are not in the
main belt (such as Apollo or Trojan asteroids), 0.7\% are comets,
0.2\% are Kuiper-belt objects, and the remaining 0.4\% is composed of
a plethora of miscellaneous objects, such as planetary satellites and
centaurs \citep{SolarSystemObjects}.  These proportions, however, are
not representative of the actual small-body populations because there
is considerable observational bias towards the closer main-belt
asteroids in comparison with more distant objects \citep{KBO_Population}.

\subsection{Generating a database of known impactors}

 We generated an ensemble of 330,000 KIs according to Algorithm \ref{generate-ki} to act as examples of hazardous objects.  Here virtual objects are launched from future positions of Earth's surface and then integrated backward in time to the present era. The idea is that the virtual objects' trajectories would be similar to that of an asteroid observed in the present that would strike the Earth or come very close to it at some point in the future. \footnote{An object, for example, that is launched from the Solar System at the year 2318, and is then integrated backwards in time 300 years, would create an example of a present day asteroid that would strike the Earth in 300 years after the velocity vectors are negated to account for the time reversal. As explained in Section \ref{SEC:arch}, the asteroids are not guaranteed to collide with Earth due to the finite precision of the integrations.} 

\begin{algorithm}[h]
\caption{KI generation algorithm. Here, $T_0$ is the earliest Solar System orientation, $T_1$ is the latest orientation, $n$ is the number of KIs, and $\Delta T=(T_1-T-0)/n$}\label{generate-ki}
\begin{algorithmic}[1]
\State{T=$[T_0, T_0 + \Delta t, T_0 + 2 \Delta t, ..., T_0 + (n-1) \Delta t, T_1]$}
        \For{each $\tau$ in $T$}
            \State \parbox[t]{\dimexpr\textwidth-\leftmargin-\labelsep-\labelwidth}{Initialize the Solar System's planets' velocities and \\ positions with values corresponding to epoch \textit{$\tau$}.}
            \State \parbox[t]{\dimexpr\textwidth-\leftmargin-\labelsep-\labelwidth}{Launch a virtual object perpendicularly from Earth's \\ surface with a velocity magnitude randomly drawn \\ from an even distribution between 15 and 45km/s.}
            \State \parbox[t]{\dimexpr\textwidth-\leftmargin-\labelsep-\labelwidth}{Integrate the object backward in time along with all \\ other Solar System objects until the present epoch.}
            \State \parbox[t]{\dimexpr\textwidth-\leftmargin-\labelsep-\labelwidth}{If the object has left the Solar System or spun into the  \\ sun, discard it and rerun the simulation.} 
            
\EndFor
\end{algorithmic}
\end{algorithm}

The future launch dates, defined by the orientation of the Solar System, are evenly distributed between 300 and 20,000 years in the future, which correspond to $T_0$ and $T_1$ values of 2318 and 22018, respectively. The launching velocities are selected to bracket the Earth's and Solar System's escape speeds of 11.2 and 42.5km/s, respectively. We deliberately did not attempt to mimic the observed asteroid impact velocities to allow the neural network to learn from the full range of parameters, rather than just based on a hand-selected subsample.

\section{Results}
\label{SEC:Results}

\subsection{Identifying Earth-impacting asteroids}

The training of the network led to the positive identification of
95.25\% of the KIs that were not part of the training and 90.99\% of the PHOs as PIs. Additionally, 1.94\% of the observed objects that were not classified as PHOs were identified as PIs. The high fraction of correctly identified KIs indicates
that HOI positively recognizes most objects that are constructed
to strike Earth.  This result is not unexpected because HOI was
specifically tuned to identify artificial KI objects. A more
meaningful metric of performance is the percentage of PHOs
identified. Although 9.01\% PHOs were not classified as potential
impactors, HOI is approximately 47 (90.99/1.94) times more likely to
select a PHO over some other observed object.

\begin{figure}[h]
        \hspace*{-0.35cm}
        \includegraphics[width=83mm]{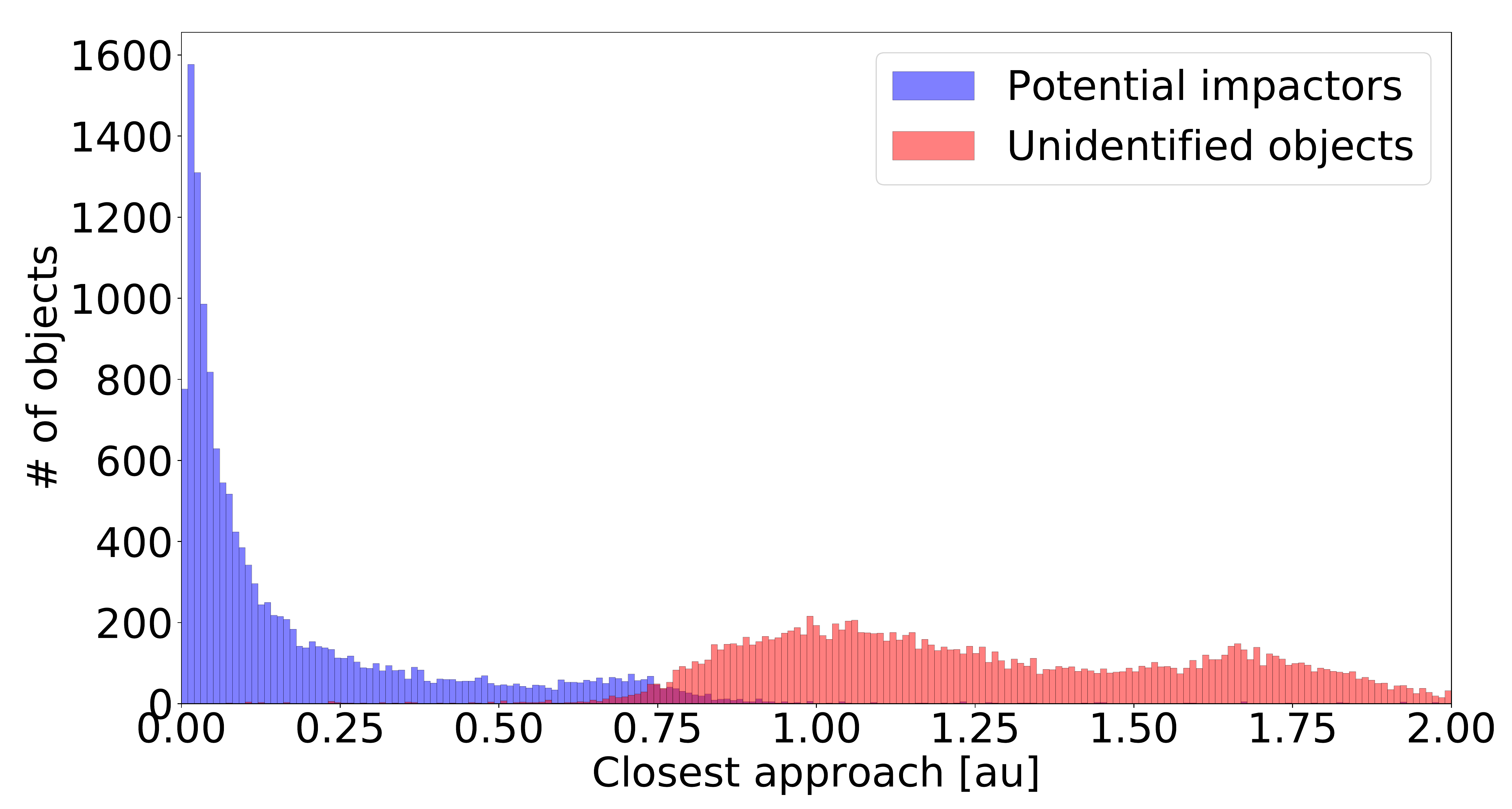}
        \centering
        \caption{\label{FIG:Closeness_Histogram} Closest approaches to Earth achieved in the next 1000 years for all the observed PIs and an equal number of randomly selected UOs. 108 PIs and 884
          UOs are not plotted because their closest
          approaches exceeded the x-axis limits of 2 \,au. Every object that reach Earth within 0.01\,au and 99.9\% of
          objects within 0.05\,au are identified by HOI as PIs.
}
\end{figure}

To further evaluate the effectiveness of HOI, we performed simulations
to compare the closest Earth approaches of PIs and UOs. To run these simulations, we began by loading the positions and velocities of the asteroids and other Solar System objects corresponding to January 1, 2018. We then integrated all of the bodies forward in time for a thousand years while saving the closest approach that the asteroids made relative to Earth. The trajectories of all the 14,680
observed PIs and an equal number of randomly selected UO asteroids were computed. The distributions of the closest Earth approaches achieved during these simulations are plotted in Fig. \ref{FIG:Closeness_Histogram}.

\begin{figure}[h]
    \hspace*{-0.44cm}
        \includegraphics[width=85mm]{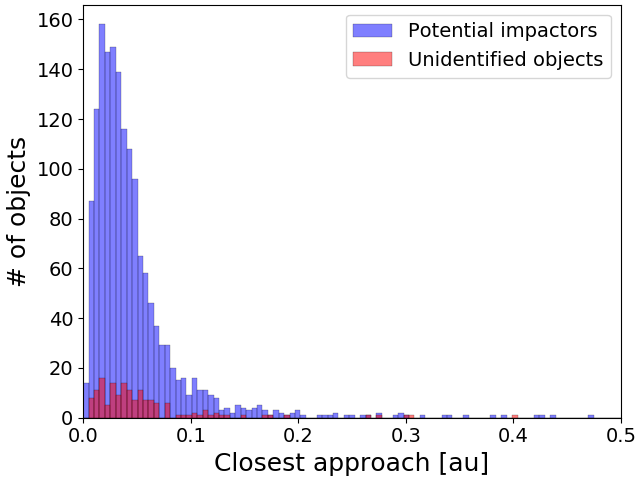}
        \centering
        \caption{\label{FIG:Closeness_PHOs} Closest approach
          distances to Earth reached for PHOs in the coming 1000 years.}
\end{figure}

To investigate why HOI only identified approximately nine-tenths of
PHOs as PIs, the thousand-year integrations described above were additionally performed
for all PHOs. We present in Fig.\,
\ref{FIG:Closeness_PHOs} the distributions of these
closest approaches.  The distributions of identified PHOs and
unidentified PHOs are similar, therefore the fraction of PHOs identified as PIs could be used as a measure of the network's
performance.  Additionally, all objects that did not approach Earth
within at least 0.5\,au could be considered misclassified PIs. This
cut-off is not arbitrary but based, rather, on the minimum distance
achieved by approximately 99.7\%, or $3\sigma$, of PHOs. In the case of HOI,
12.2\% of the PIs are outside of this threshold and are therefore
considered misclassified. The root of this misclassification  likely stems
from the approximations made in the labeling schemes described in Section \ref{SEC:arch}.

A total of $13,258$ asteroids identified by HOI as KIs are not listed by
NASA as PHOs.  In our thousand-year integrations, $4472$ of these objects
approached within 0.05\,au of Earth while $2015$ approached within 0.02\,au. In Table.\,\ref{TAB:Short_List} we present a short
list of 11 notable asteroids with absolute magnitudes of
less than 22, data arcs of less than 31 days, and closest approaches less than 0.02\,au.

\begin{table}[]
\begin{tabular}{lcrlr}
\hline \hline
\bf{Designation} & \bf{CA} & \bf{$t_{\rm CA}$} & \bf{H} & \bf{arc} \\
                 & [au]    & [Year] & [mag] & [day] \\
\hline 
2005 RV24 & 0.020 &  Feb. 2374 & 20.60 & 28 \\
2008 UV99 & 0.013 &  April 2332 & 20.03 & 1 \\
2011 BU10 & 0.006 &  April 2920 & 21.30 & 18 \\ 
2011 HH1 & 0.012 &  July 2923 & 21.7 & 13 \\ 
2011 WC44 & 0.018 & Feb. 2679 & 20.5 & 31 \\ 
2013 AG76 & 0.013 &  Dec. 2638 & 20.3 & 24 \\ 
2014 GL35 & 0.018 & July 2556 & 20.6 & 23 \\ 
2014 TW57 & 0.017 & Sept. 2165 & 20.1 & 24 \\ 
2014 WD365 & 0.017 & Sept. 2735 & 19.7 & 5 \\
2017 DQ36 & 0.013 &  Dec. 2131 & 19.3 & 29 \\ 
2017 JE3 & 0.016 & July 2741 & 21.9 & 23 \\
\hline \hline
\vspace{0.5cm}
\end{tabular}
\label{TAB:Short_List}
\caption{Potential impactor shortlist: relatively large minor
  bodies with a short data arcs that were identified as PIs by HOI but are not considered PHOs. Along with their closest approaches (CA)
  in au, the month and year that their closest approaches occurred
  ($t_{\rm CA}$), their absolute magnitudes (H), and their data arc
  lengths in days (arc) are tabulated.}
\end{table}

The absolute magnitude threshold of 22 was chosen so that only
asteroids that have the potential of causing regional devastation
unprecedented in human history would make the shortlist. Assuming a geometric
albedo between 0.05 and 0.25 and a spherical shape, objects with an absolute magnitude of
22 are estimated to have diameters between from 100\,m to 236\,m. For perspective, Tunguska object which flattened 2,000 square kilometers of forest in
Siberia was estimated to have a diameter of between 50-80\,m \citep{Tunguska}.
The month long data-arc limit is selected because the Monte-Carlo
method adopted by NASA is particularly ill-suited for calculating the
impact probabilities of such uncertain orbits.  As a consequence,
these objects are the most likely to be overlooked as PHOs.

\subsection{Comparing various populations of object}

\begin{figure*}[t!]
    \hspace*{-0.40cm}
        \includegraphics[width=170mm]{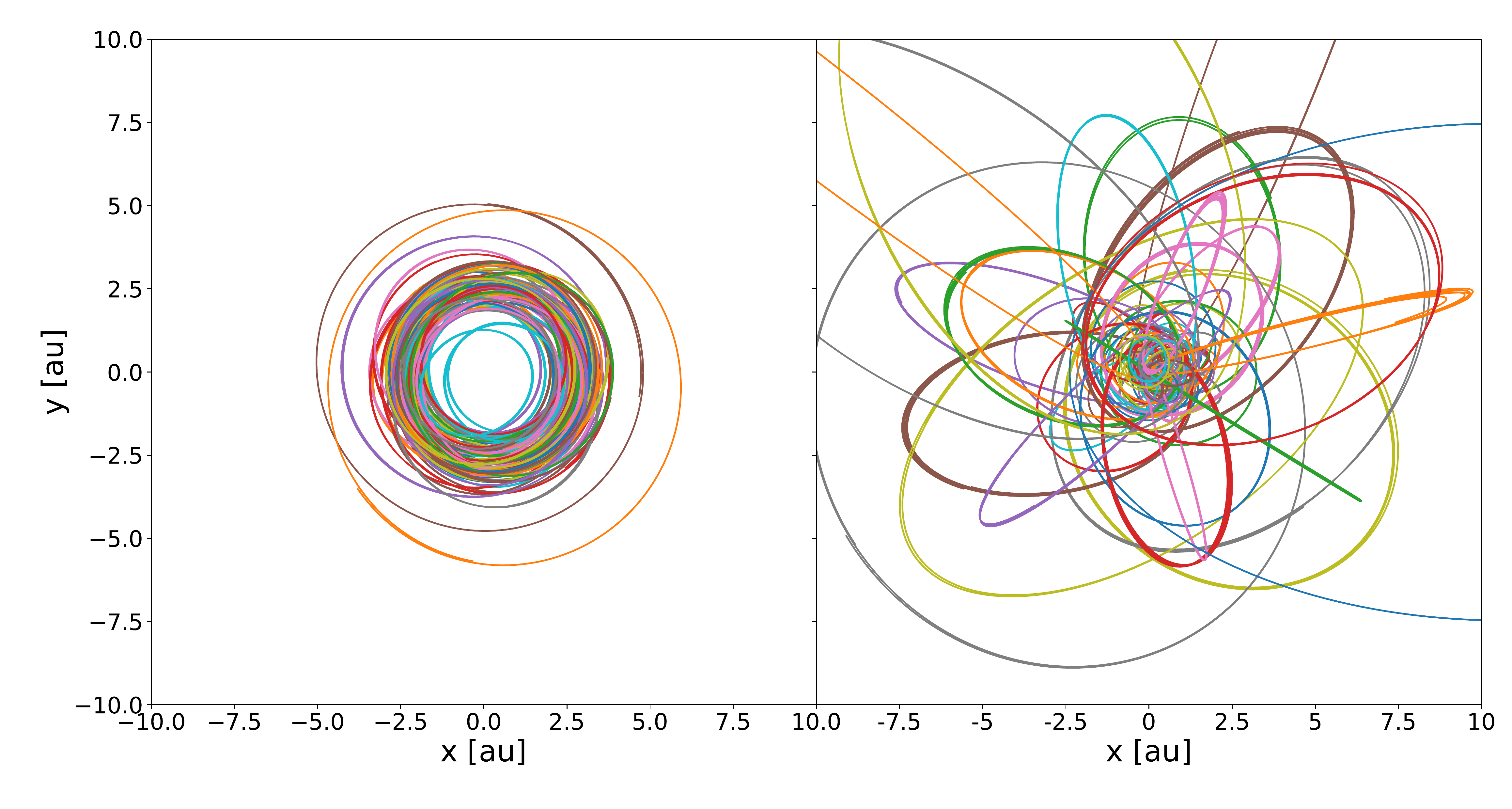}
        \centering
        \caption{\label{FIG:Object_Trajectories} Illustration of the
          difference between the trajectories of observed objects
          (left) and KIs (right). The observed objects tend to have circular orbits
          which lie in the orbital plane of Earth around the Sun, whereas
          the KIs exhibit a much broader distribution in
          eccentricity and inclination. These characteristics, however,
          are not mutually exclusive and could be one the root causes
          of HOI's imperfect classification.}
\end{figure*}

The characteristics of the simulated KIs and the observed objects are
compared to better understand how HOI differentiates between the two
populations. In Fig. \ref{FIG:Object_Trajectories} we present 100
trajectories of observed objects and KIs.

There are profound differences between the orbital elements of the two
distinct populations of objects. Our artificial population of objects
launched from Earth tend to have highly eccentric and inclined orbits,
whereas the observed objects tend to have circular orbits confined
near the ecliptic plane. For the observed objects, the orbital plane
is essentially empty within approximately 2\,au of the Sun, while for
the KIs this is the most densely occupied space. This object
distribution should be expected considering that all the KIs
were generated $1\pm0.017$\,au away from the Sun along the Earth's
orbit and that the integration times were not sufficiently long enough to allow considerable outward migration of the objects.

\begin{figure*}[t!]
        \hspace*{-0.50cm}
        \includegraphics[width=170mm]{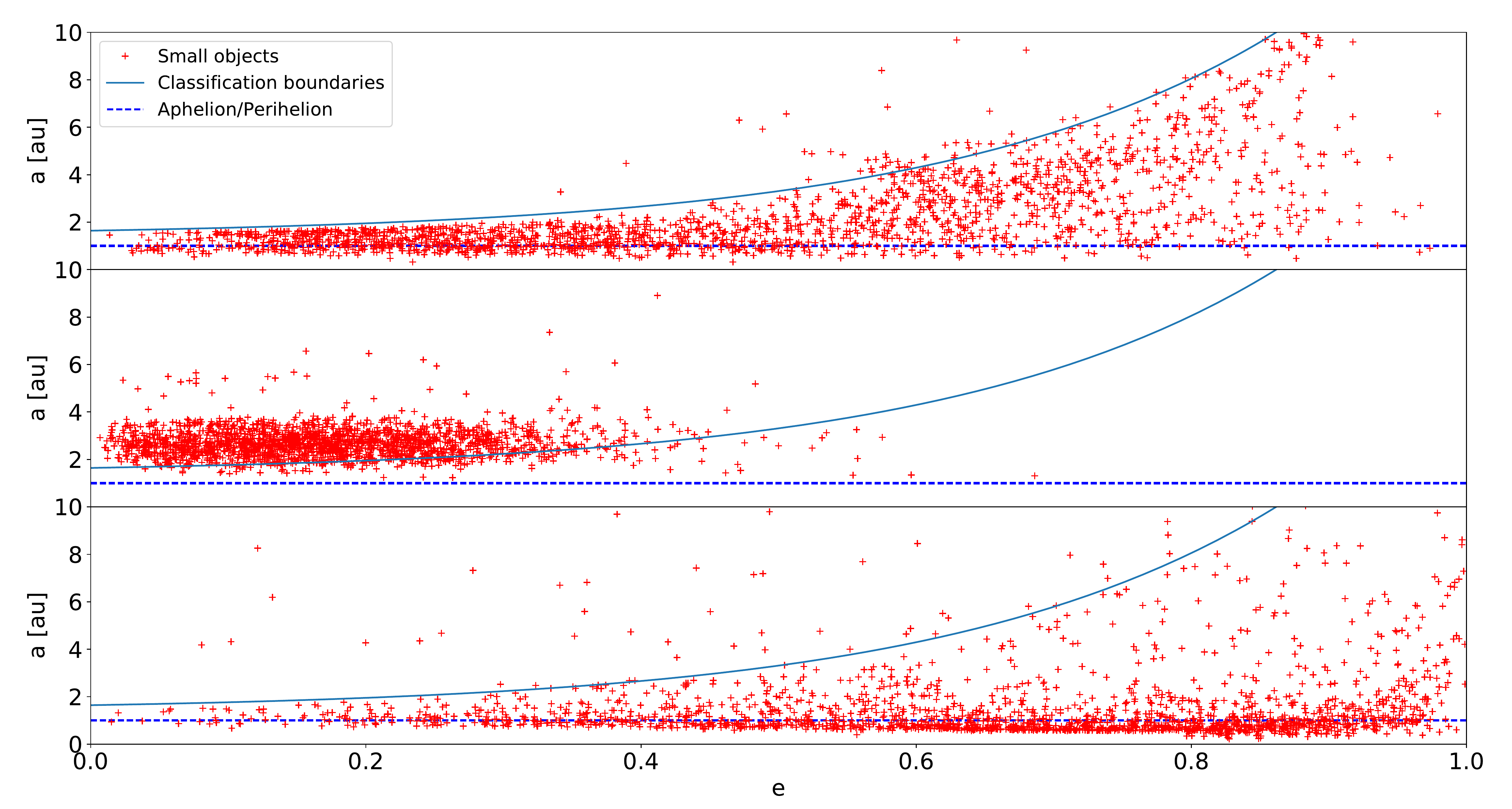}
        \centering
        \caption{\label{FIG:AvsE} Plots the semi-major axis versus the eccentricity for 2,000 PI, UO and KI objects, respectively, from top to bottom. The dotted blue lines represent the aphelion and perihelion distances of Earth's orbit and the teal curves represent the ``classification boundary'' where objects below are likely to be classified as PIs and those above are likely to be classified as benign.}
        
\end{figure*}

The \textit{a} versus \textit{e} ratio is an important factor in an object's identification, as illustrated in Fig. \ref{FIG:AvsE}. A curve is drawn to highlight an apparent ``classification boundary'', which is above 95.2\% of PI and below 90.3\% of unidentified observed objects. Although the boundary is an indicator of an object's potential classification, it is not definite, which is understandable considering that HOI takes five orbital elements as input for each object instead of just the \textit{a} and \textit{e} orbital elements.

\section{Conclusions}
\label{SEC:Conclusions}

We designed, constructed, and trained a fairly simple neural network aimed at
classifying asteroids with the potential to  impact the Earth over the coming
$20,000$ years. Our method takes the observed orbital elements as
input and provides a classifier for the expectation value for the object's
striking Earth.

The network was able pick out 95.25\% of the KIs
when mixed into a set of observed asteroids which are not expected to
strike Earth. When applied to the entire population of observed
asteroids, the network was able to identify approximately nine-tenths of the
asteroids identified by NASA as PIs and along with virtually
every other observed asteroid that approached within 0.05\,au of Earth.  We generated
a short list of network identified PIs which
NASA does not label as PHOs, mainly because the observed
orbital elements are so uncertain that NASA's Monte Carlo approach to
determine their Earth-striking probability fails.
The network classifies an object as a PI or UO within $0.25$
milliseconds, which is negligible compared to the time required for
the Monte-Carlo method employed by NASA.
 
Follow-up calculations over a time-span of 1000 years revealed that
12.2\% of the PIs identified by the network did not
come within 0.5\,au of Earth. This may imply that thee asteroids pose
no direct threat on the time scale considered. Integrating their
orbits for a longer time-frame, however, this is impractical because of the
large uncertainty in their orbital elements and the relatively small
Lyapunov timescale for these objects.

We look forward to improving the network's classification accuracy.  The
network, as we show in Fig.\,\ref{FIG:HOI_Design}, is the result
of a great deal of experimentation in network depth, width, and  (sub)selection
input parameters. It is possible that the structure preserving mimetic
architectures motivated by the underlying Keplerian topology of the
orbits could allow us to achieve a higher quality of prediction accuracy but this still
requires a considerable degree of further experimentation.  Another improvement could be
carried out by considering a stricter labeling scheme in which some
probability statistics for impacting the Earth could be taken into
account.

\begin{acknowledgements}
  
  We thank the Microsoft Cooperation for access to the Azure cloud on
  which many of the calculations presented here are performed.  John D. Hefele
  thanks Sander van den Hoven for his mentoring during his internship
  at Microsoft Amsterdam. This work was supported by the Netherlands
  Research School for Astronomy (NOVA), NWO (grant \# 621.016.701
  [LGM-II]).

\end{acknowledgements}
\bibliographystyle{aa}
\bibliography{bibliography}

\end{document}